\input{aipcheck}

\documentclass[
    ,final            
  ]
  {aipproc}

\layoutstyle{6x9}

\begin{document}

\title{New analysis of the common nuclear dependence of the EMC effect and short-range correlations}



\author{Nadia Fomin}{
  address={Los Alamos National Laboratory, Los Alamos, NM, 87545, USA}
}

\begin{abstract}
The strong repulsive core of the nucleon-nucleon (NN) interaction at short distances prevents nucleons from becoming close to each other. This gives rise to high-momentum nucleons in the nucleus that cannot be explained in the context of the mean field and are commonly called short-range correlations (SRCs). They are responsible for the strength seen in momentum distribution tails seen in all nuclei, and we can obtain a relative measure of SRCs via cross section ratios to light nuclei. Recent inclusive scattering data from Jefferson Lab have allowed a precise determination of the A-dependence of SRCs in nuclei and suggests that, like the EMC effect, it is especially sensitive to the nuclear local density. These new results, as well as a new analysis of the relationship between SRCs and the EMC effect, will be presented and discussed.
\end{abstract}

\maketitle


\section{EMC effect}

The EMC effect was observed~\cite{aubert83} in 1983 when the ratio of Fe/D cross sections was found to deviate from unity in the Deep Inelastic Scattering (DIS) region.  This depletion of strength for heavy nuclei meant that the nuclear structure functions are not simply the sum of neutron and proton structure functions, but are instead modified by the nuclear medium. 

\begin{figure}[ht]
\includegraphics[height=.3\textheight,angle=270]{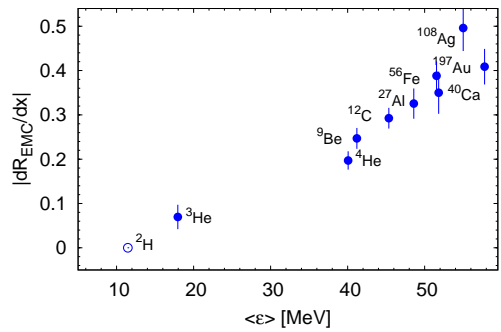}
\label{fig1}
\caption{$|dR_{EMC}/dx|$ as a function of average separation energy, $\epsilon$, from Ref.~\cite{kulagin2006}}.
\end{figure}

 Detailed studies of the EMC effect have since been performed~\cite{gomez94, seely09}, confirming a universal shape and a magnitude that was seen to scale with $A$ or average nuclear density, with the exception of $^9$Be, which acts like a denser nucleus.  
This recent observation has renewed interest in understanding what drives the EMC effect, and we have tested several quantities for a linear relationship.  The average nucleon removal energy, $\epsilon$ yielded a  very striking result, seen in Fig.~\ref{fig1}.
However, the origin of the EMC effect remains a mystery for now.

\section{Short-range correlations}
Quasielastic scattering is used to probe high momentum nucleons in short-range correlations through ratios of $A/D$ cross sections.  The inclusive scattering cross section can be written in terms of scattering from $j$-nucleon correlations via $\sigma(x,Q^2) = \sum_{j=1}^{A} A\frac{1}{j}a_j(A) \sigma_j(x,Q^2)$
%
%
where $\sigma_j(x,Q^2)=0$ at $x>j$ and the $a_j(A)$'s are proportional to the
probabilities of finding a nucleon in a $j$--nucleon correlation.  Focussing on 2N correlations, the expectation is that the high momentum tails for $A>$2 will be rescaled versions of the deuteron, meaning that the measured ratios should exhibit scaling plateaus for $x>$1.4, where mean field contributions ($k<k_{fermi}$) become neglible.  These plateaus have been observed in numerous experiments~\cite{frankfurt93,egiyan03,egiyan06,fomin2012}, with the magnitude of the plateaus giving a relative measure of the probability of 2N correlations in a nucleus to that in the deuteron.

\section{Testing the SRC-EMC relationship}
A linear relationship was recently observed~\cite{weinstein2010rt} between the slope of the EMC effect and the SRC plateau and re-examined~\cite{Hen:2012fm,Arrington:2012ax} using new data~\cite{fomin2012}.  There are two hypotheses put forth to explain this.  The first~\cite{weinstein2010rt} suggests that the EMC effect is driven by the high virtuality (HV) of the nucleons, which is reflected by $a_2$, the ratio of $A/D$ cross sections in the high-momentum region.  The second proposes that ``local density'' (LD) is at the root of the relationship, supported by the fact that $^9$Be, whose average density is low, can be described as two tight alpha-like clusters~\cite{seely09}, giving rise to an EMC effect and SRC ratio comparable to those of carbon or $^4$He.  To test this hypothesis, we need a quantity that represents ``local density''.  We correct the raw  $A/D$ cross section ratio for the center of mass motion of the correlated pair for $A>$2, removing an enancement of the high-momentum tail.  This corrected ratio, $R_{2N}$, then represents the number of $np$ correlated pairs, which make up over 90\% of short-range configurations~\cite{subedi08}.  However, the EMC effect samples all nuclei, and if it's driven by ``local density'', it will be sensitive to \textit{all} short-range configurations.  Therefore, before testing the LD hypothesis, we scale $R_{2N}$ by $N_{tot}/N_{iso}$, the ratio of all NN pairs to $np$ pairs only. 

\begin{figure}[h!]
\label{fig2}
\caption{Slope of the EMC effect (y-axis) vs $a_2$ (left) and $R_{2N}$, scaled by $N_{tot}/N_{iso}$.  The red dotted lines are 2 parameter unconstrained fits, while the solid black lines are 1 parameter fully constrained fits.}
\includegraphics[height=.3\textheight,angle=270]{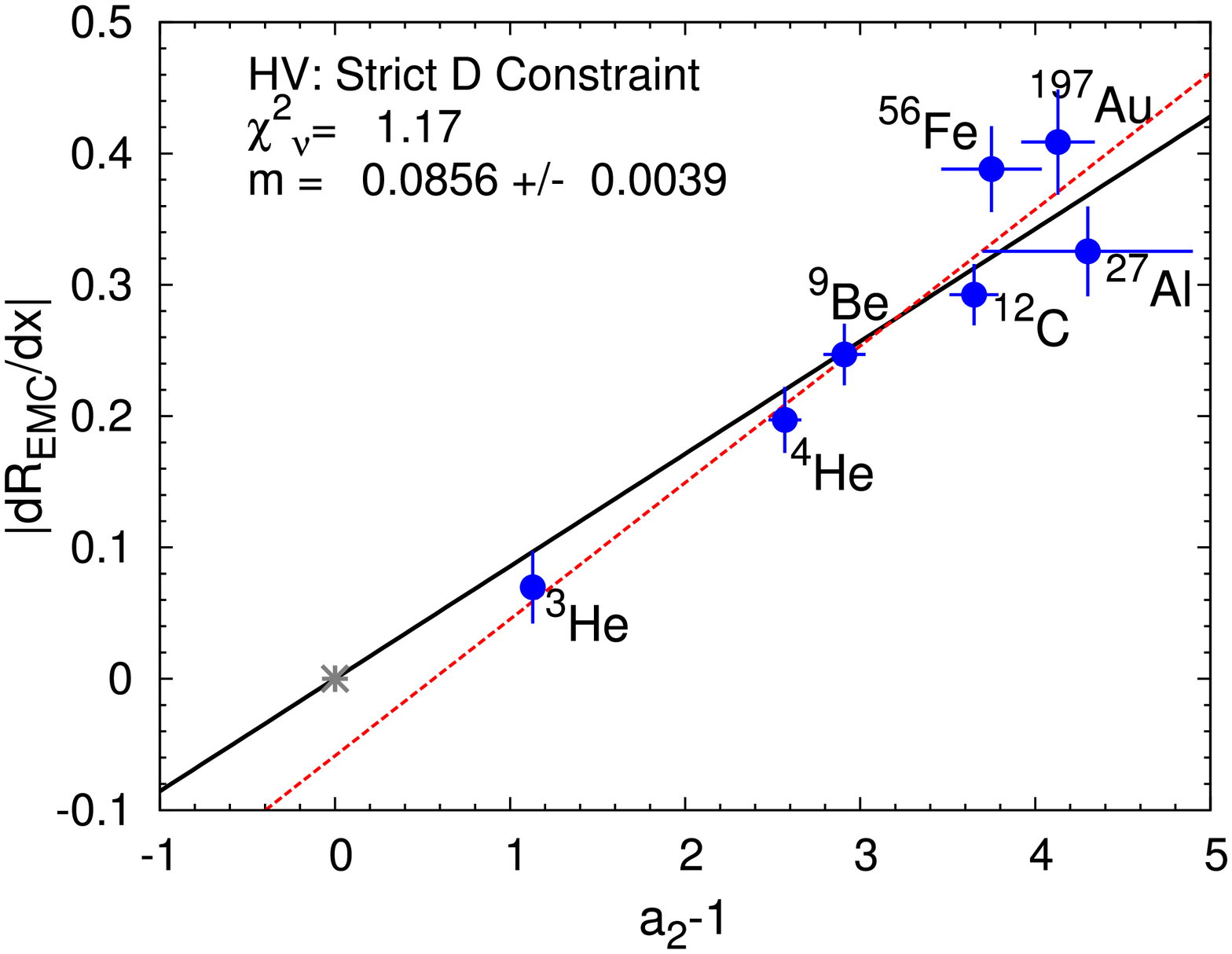}
\includegraphics[height=.3\textheight,angle=270]{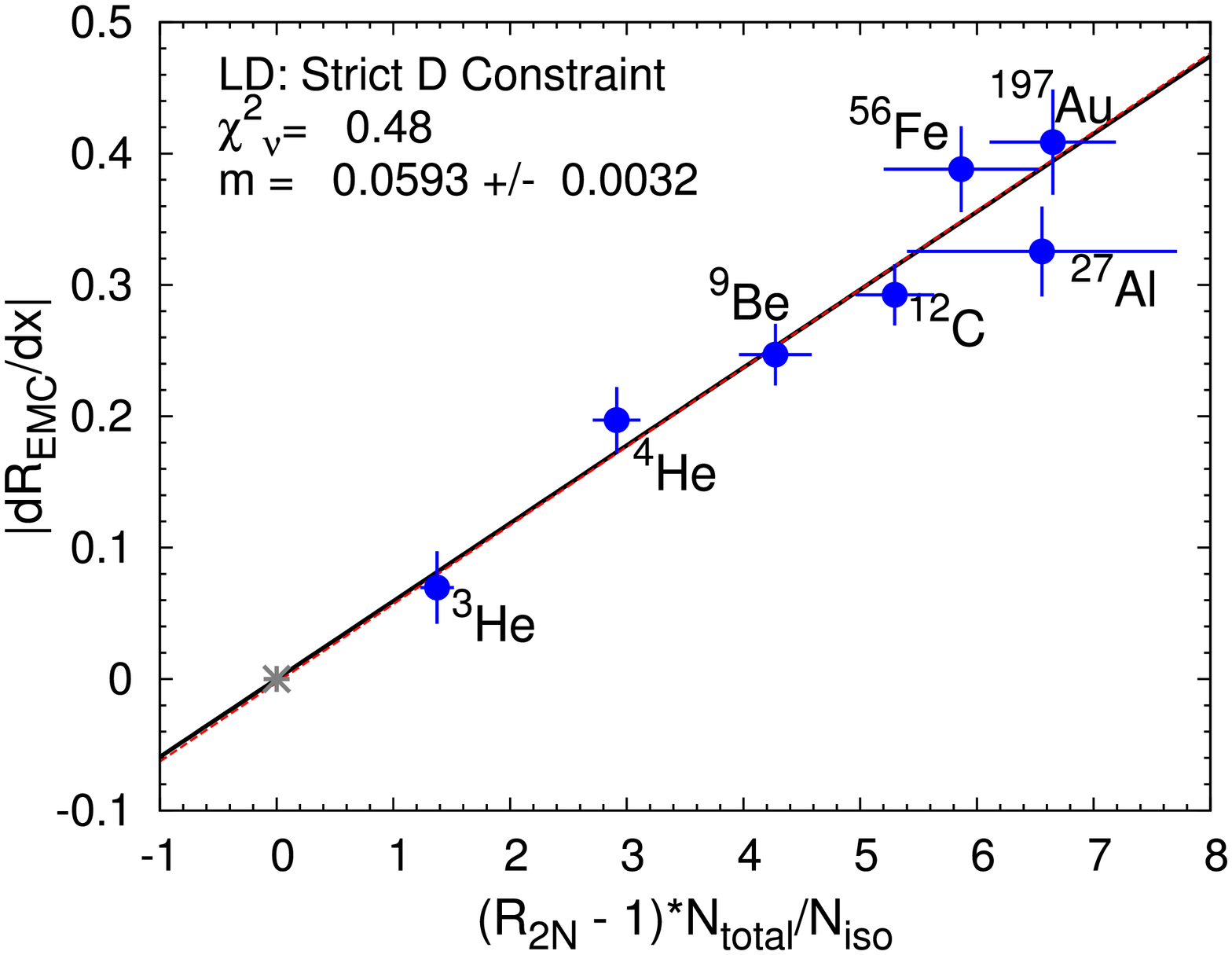}
\end{figure}

This procedure was repeated with a strict deuteron constraint (no offset in the fit) as well as a more reasonable deuteron constraint, with errorbars derived from EMC and SRC measurements.  The results of the linear tests with 2 free parameters as well as the strict deuteron fit using both hypotheses can be seen in Fig.~\ref{fig2}.  All the results are summarized in Tab.~\ref{tab1}, including slopes, $\chi^2/_{\nu}$ values, EMC effect for the deuteron (where applicable), as well as the IMC effect for the deuteron, defined in~\cite{weinstein2010rt} to be the medium modification in a nucleus compared to the sum of a free proton and neutron.

\begin{table}
\vspace*{0.25in}
\begin{tabular*}{.95\textwidth}{@{\extracolsep{\fill}}|c|c|c|c|}
\hline
Test & $\chi^2_\nu$ & EMC(D) & IMC(D)\\
\hline
\hline

HV        &~0.91       & -0.0587$\pm$0.037 & 0.1040$\pm$0.012 \\
HV-0      &~1.17       & --                & 0.0856$\pm$0.004 \\
HV-D      &~1.14       & -0.0041$\pm$0.010 & 0.0869$\pm$0.005 \\
LD        &~0.57 (0.74)& -0.0028$\pm$0.033 & 0.0599$\pm$0.008 \\
LD-0      &~0.48 (0.61)& --                & 0.0593$\pm$0.003 \\ 
LD-D      &~0.48 (0.61)& -0.0002$\pm$0.010 & 0.0593$\pm$0.004 \\ 
\hline
\end{tabular*}

\caption{Summary of linear fits of EMC effect vs $R_{2N}$ or $a_2$, and extrapolations to the slopes of the EMC effect for the deuteron, EMC(D), and IMC effect for the deuteron, IMC(D).''-0'' denotes a 1-parameter fit, forcing the line to go through zero, corresponding to no EMC effect for the deuteron. ``-D'' denotes a two parameter fit  including a realistic deuteron constraint described in the text.  Number in parentheses of the $\chi^2_\nu$ column includes the result of fitting with smaller fractional errors from $a_{2}$.}
\label{tab1}
\end{table}

Current data do not clearly favor one hypothesis over the other.  To shed more light on this, there are approved experiments at Jefferson Lab to run after the 12 GeV upgrade is complete to measure the EMC effect~\cite{E1210008} as well as SRCs~\cite{E1206105} in many additional nuclei.

\bibliographystyle{aipproc}   

\bibliography{330_fomin}

\IfFileExists{\jobname.bbl}{}
 {\typeout{}
  \typeout{******************************************}
  \typeout{** Please run "bibtex \jobname" to optain}
  \typeout{** the bibliography and then re-run LaTeX}
  \typeout{** twice to fix the references!}
  \typeout{******************************************}
  \typeout{}
 }

\end{document}